\documentclass[10pt]{article}
\usepackage{OSAmeet}
\def\bfR{{\bf R}}
\def\bfZ{{\bf Z}}
\def\bfr{{\bf r}}
\def\defi{\stackrel{\rm def}{=}}
\begin{document}
\title{On the Capacity of Nonlinear Fiber Channels}
\author{Haiqing Wei$^*$ and David V. Plant}
\address{Department of Electrical and Computer Engineering\\
McGill University, Montreal, Canada H3A-2A6}
\email{$^*$hwei1@po-box.mcgill.ca} \pagestyle{plain}
\begin{abstract}
The nonlinearity of a transmission fiber may be compensated by a
specialty fiber and an optical phase conjugator. Such combination
may be used to pre-distort signals before each fiber span so to
linearize an entire transmission line.
\end{abstract}

A fiber-optic transmission line is a nonlinear channel due to the
material nonlinear effects. Fiber nonlinearity has become one of
the major limiting factors in modern optical transmission systems
\cite{Forghieri97,Mitra01}. Distributed Raman amplification and
various return-to-zero (RZ) modulation formats may be employed to
reduce the nonlinear impairments, but merely to a limited extent.
It is known that the nonlinearity of one fiber line may be
compensated by that of another with the help of optical phase
conjugation (OPC). However, all previous proposals and
demonstrations
\cite{Pepper80,Watanabe94,Watanabe96,Grandpierre99,Brener00} work
partially in fighting the fiber nonlinearity. They either are
specialized to only one aspect of the nonlinear effects, or fail
to work in the presence of dispersion slope or higher order
dispersion effects. Still open to date is an important question:
does fiber nonlinearity really impose a fundamental limit to the
channel capacity? This paper introduces the notion of scaled
nonlinearity, which together with the application of OPC, provides
a negative answer to the above question.

Consider an optical fiber stretching from $z=-L$ to $z=0$, along
which the optical birefringence is either vanishingly weak to
avoid the effect of polarization mode dispersion (PMD), or
sufficiently strong to render the fiber polarization maintaining.
A set of optical signals are wavelength-division multiplexed (WDM)
and injected into the fiber. For simplicity, it is assumed that
all the signals are co-linearly polarized, and coupled into one
polarization eigen state when the fiber is polarization
maintaining. The signals may be represented by a sum of scalars
$E(\bfr,t)=F(x,y)\sum_nA_n(z,t)\exp\left[i\int^z\beta(\zeta,
\omega_n)d\zeta-i\omega_nt\right]$, where $F(x,y)$ is the
transverse modal function, $\forall n\in\bfZ$,
$\omega_n=\omega_0+n\Delta$ and $A_n$ are the center frequency and
the signal envelope of the $n$th WDM channel respectively,
$\Delta>0$ is the channel spacing, $\beta(z,\omega)$ is the
function of propagation constant vs. optical frequency $\omega$ at
the position $z$ in the fiber. Define $\beta^{(k)}(z,\omega)\defi
\frac{\partial^k}{\partial\omega^k}\beta(z,\omega)$, $\forall
k>0$. The dynamics of signal propagation may be described by a
group of coupled partial differential equations
\cite{Shen84,Agrawal95},
\begin{equation}
\frac{\partial A_n}{\partial
z}-i\beta_{1n}\left(z,i\frac{\partial}{\partial
t}\right)A_n+\frac{\alpha_n(z)}{2}A_n=i\gamma(z)\sum_k\sum_l
A_kA_lA_{k+l-n}^*\exp[i\theta_{kln}(z)]+\sum_mg_{mn}(z)|A_m|^2A_n,
\label{dAn1}
\end{equation}
$\forall n\in\bfZ$, $-L\le z\le 0$, where $\gamma$ is the Kerr
nonlinear coefficient, $\alpha_n$ is the attenuation coefficient
around $\omega_n$, $g_{mn}$ is the Raman coupling coefficient from
the $m$th to the $n$th channels,
$\theta_{kln}(z)\defi\int^z[\beta(\zeta,\omega_k)+
\beta(\zeta,\omega_l)-\beta(\zeta,\omega_{k+l-n})-
\beta(\zeta,\omega_n)]d\zeta$ denotes the phase mismatch among the
mixing waves, and $\beta_{1n}$ is a functional operator defined as
$\beta_{1n}(z,i\frac{\partial}{\partial
t})\defi\sum_{k=1}^{+\infty}\frac{1}{k!}\beta^{(k)}(z,\omega_n)
\left(i\frac{\partial}{\partial t}\right)^k$. Using the frame
transformation $(z,t)\rightarrow\left(z,t+\int^z\beta^{(1)}(\zeta,
\omega_0)d\zeta\right)$, (\ref{dAn1}) may be rewritten as,
\begin{equation}
\frac{\partial A_n}{\partial
z}-i\beta_{2n}\left(z,i\frac{\partial}{\partial
t}\right)A_n+\frac{\alpha_n(z)}{2}A_n=i\gamma(z)\sum_k\sum_l
A_kA_lA_{k+l-n}^*\exp[i\theta_{kln}(z)]+\sum_mg_{mn}(z)|A_m|^2A_n,
\label{dAn2}
\end{equation}
$\forall n\in\bfZ$, $-L\le z\le 0$, with
$\beta_{2n}\left(z,i\frac{\partial}{\partial t}\right)
\defi\beta_{1n}\left(z,i\frac{\partial}{\partial t}\right)
-\beta^{(1)}(z,\omega_0)\left(i\frac{\partial}{\partial
t}\right)$. This set of equations completely determine the
propagation dynamics of the optical signals. In case the signals
are not co-linearly polarized, the mathematical description should
be slightly modified to deal with the complication. However, the
same physics remains to govern the nonlinear signal propagation in
optical fibers, still valid are the method of nonlinearity
compensation described below and the conclusion about the capacity
of nonlinear channels.

When the signals become intense, the nonlinear interaction as
evidenced in (\ref{dAn2}) can badly distort them and make the
carried information difficult to retrieve. Unlike the case of
linear channels \cite{Shannon48}, simply raising the signal power
may not necessarily increase the capacity of a nonlinear channel.
Nevertheless, the nonlinear interaction among the signals is
deterministic in nature that can be undone in principle. The
question is how to implement a physical device which undoes the
distortion. Suppose there is a specialty fiber stretching from
$z=0$ to $z=L/R$, $R>0$ being a constant, in which $q(z,\omega)$
is the optical propagation constant. Let a set of WDM signals
$E'(\bfr,t)=F'(x,y)\sum_nA'_n(z,t)\exp\left[i\int^z
q(\zeta,\omega'_n)d\zeta-i\omega'_nt\right]$ propagate in the
specialty fiber, where $\forall n\in\bfZ$,
$\omega'_n=\omega'_0+n\Delta$, $\omega'_0$ may differ from
$\omega_0$. As in the previous mathematical treatment, define a
functional operator $q_{2n}(z,i\frac{\partial}{\partial
t})\defi\sum_{k=1}^{+\infty}\frac{1}{k!}q^{(k)}(z,\omega'_n)
\left(i\frac{\partial}{\partial t}\right)^k-
q^{(1)}(z,\omega'_0)\left(i\frac{\partial}{\partial t}\right)$,
with $q^{(k)}(z,\omega)\defi\frac{\partial^k}{\partial\omega^k}
q(z,\omega)$, and let $\alpha'_n$, $\gamma'$, $\theta'_{kln}$,
$g'_{mn}$ denote the linear and nonlinear parameters associated
with the specialty fiber and the new set of WDM signals, then the
propagation dynamics is governed by a similar group of equations,
\begin{equation}
\frac{\partial A'_n}{\partial
z}-iq_{2n}\left(z,i\frac{\partial}{\partial
t}\right)A'_n+\frac{\alpha'_n(z)}{2}A'_n=i\gamma'(z)\sum_k
\sum_lA'_kA'_lA'^*_{k+l-n}\exp[i\theta'_{kln}(z)]+
\sum_mg'_{mn}(z)|A'_m|^2A'_n, \label{dAn3}
\end{equation}
$\forall n\in\bfZ$, $0\le z\le L/R$. If the parameters are set
according to the following rules of scaling,
\begin{eqnarray}
q^{(2)}(z,\omega'_0+\omega)&=&R\beta^{(2)}(-Rz,\omega_0-\omega),
~~\forall\omega\in\bfR,\label{beta2q}\\
\alpha'_n(z)&=&-R\alpha_{-n}(-Rz),~~\forall
n\in\bfZ,\label{alpha'}\\
\gamma'(z)&=&R\gamma(-Rz)|C|^{-2},\label{gamma'}\\
g'_{mn}(z)&=&-Rg_{-m,-n}(-Rz)|C|^{-2},~~\forall m,n\in\bfZ,
\label{gmn'}
\end{eqnarray}
$\forall z\in[0,L/R]$, $C\neq 0$ being a constant, then equations
(\ref{dAn3}) are reduced from (\ref{dAn2}) by taking the complex
conjugate, making a substitution $z\rightarrow-Rz$, and replacing
$A^*_{-n}(-Rz,t)$ by $A'_n(z,t)/C$. Mathematically, it says that
$A'_n(z,t)=CA^*_{-n}(-Rz,t)$, $\forall n\in\bfZ$, solve
(\ref{dAn3}), which govern the nonlinear propagation in the
specialty fiber. Interpreted physically, if OPC is applied after
the transmission but before the specialty fibers to convert the
signals $A_n(0,t)$, $n\in\bfZ$, into $A'_n(0,t)=CA^*_{-n}(0,t)$,
$n\in\bfZ$, then the specialty fiber will propagate the optical
signals in a reversed manner with respect to the transmission
fiber. At the end, the specialty fiber outputs signals
$A'_n(L/R,t)=CA^*_{-n}(-L,t)$, $n\in\bfZ$, which are replicas of
the initial signals before entering the transmission fiber up to
complex conjugation. The fibers and optical signals on the two
sides are said to be mirror symmetric about the OPC, although in a
scaled sense. Note that the specialty fiber would amplify light in
correspondence to the attenuation in the transmission fiber and
vice versa.

\begin{figure}[h]
\centerline{\scalebox{.5}{\includegraphics{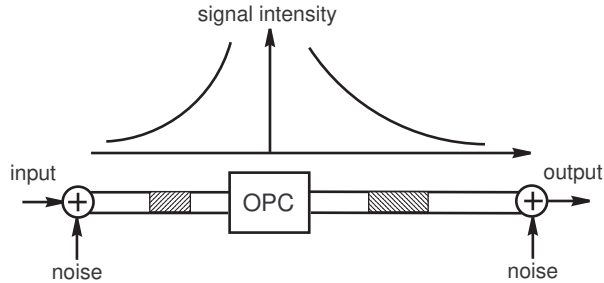}}}
\caption{\label{mirror}A specialty fiber and a transmission fiber
are in scaled mirror symmetry about the point of OPC. The shaded
fiber segments are located at $-z$ and $Rz$ respectively, $0\le
z\le L/R$, with the parameters $q^{(2)}(-z)$, $\alpha'(-z)$,
$\gamma'(-z)$, $g'(-z)$ proportional to $\beta^{(2)}(Rz)$,
$\alpha(Rz)$, $\gamma(Rz)$, $g(Rz)$ as in
(\ref{beta2q}-\ref{gmn'}).}
\end{figure}

So OPC and a specialty fiber with parameters designed according to
(\ref{beta2q}-\ref{gmn'}) could perfectly compensate the
nonlinearity of a transmission fiber, if not for the ever-existing
noise, especially that incurred when the signal amplitude is low,
destroying the mirror symmetry. A better designed link would start
with a specialty fiber that boosts the power of the optical
signals, followed by OPC, then a fiber for transmission in which
the signal power decreases, as shown in Fig.\ref{mirror}. In the
fiber locations not too far from the OPC, the signal power is
relatively high to minimize the effect of the optical noise, which
usually originates from amplified spontaneous emission (ASE) and
quantum photon statistics. However at the two ends of the link,
the effect of the optical noise could become substantial. A simple
but fairly accurate model may assume that optical noise is
incurred exclusively at the two extreme ends of the link,
dispersive and nonlinear signal propagation is the only effect of
the inner part of the link. In this model, the nonlinearity of a
segment of transmission fiber with $z_1\le z\le z_2$ is fully
compensated by the portion of the specialty fiber with $-z_2/R\le
z\le -z_1/R$, $\forall z_1,z_2\in[0,L]$. In particular, the entire
link from $z=-L/R$ to $z=L$ is equivalent to a linear channel
impaired by additive noise at the two ends. If $W$ is the total
optical bandwidth of the input WDM channels, then the OPC should
have a bandwidth wider than $W$ to cover the extra frequency
components generated through wave mixing in the specialty fiber.
With nonzero dispersion fibers, however, the extra bandwidth due
to wave mixing may hardly exceed $100$ GHz, which is often
negligible in comparison to the total bandwidth $W$ of several,
even tens of THz. Thus the linearized link may be assumed to have
the same bandwidth limit $W$ throughout, applicable to which is
Shannon's formula for channel capacity \cite{Shannon48},
$C=W\log_2(1+S/N)$. Obviously, many of such linearized links may
be cascaded to reach a longer transmission distance, and the
entire transmission line is still linear end-to-end in spite of
the nonlinearity existing locally in the fibers.

\begin{figure}[h]
\begin{center}
\begin{minipage}{78mm}
\scalebox{.37}{\includegraphics{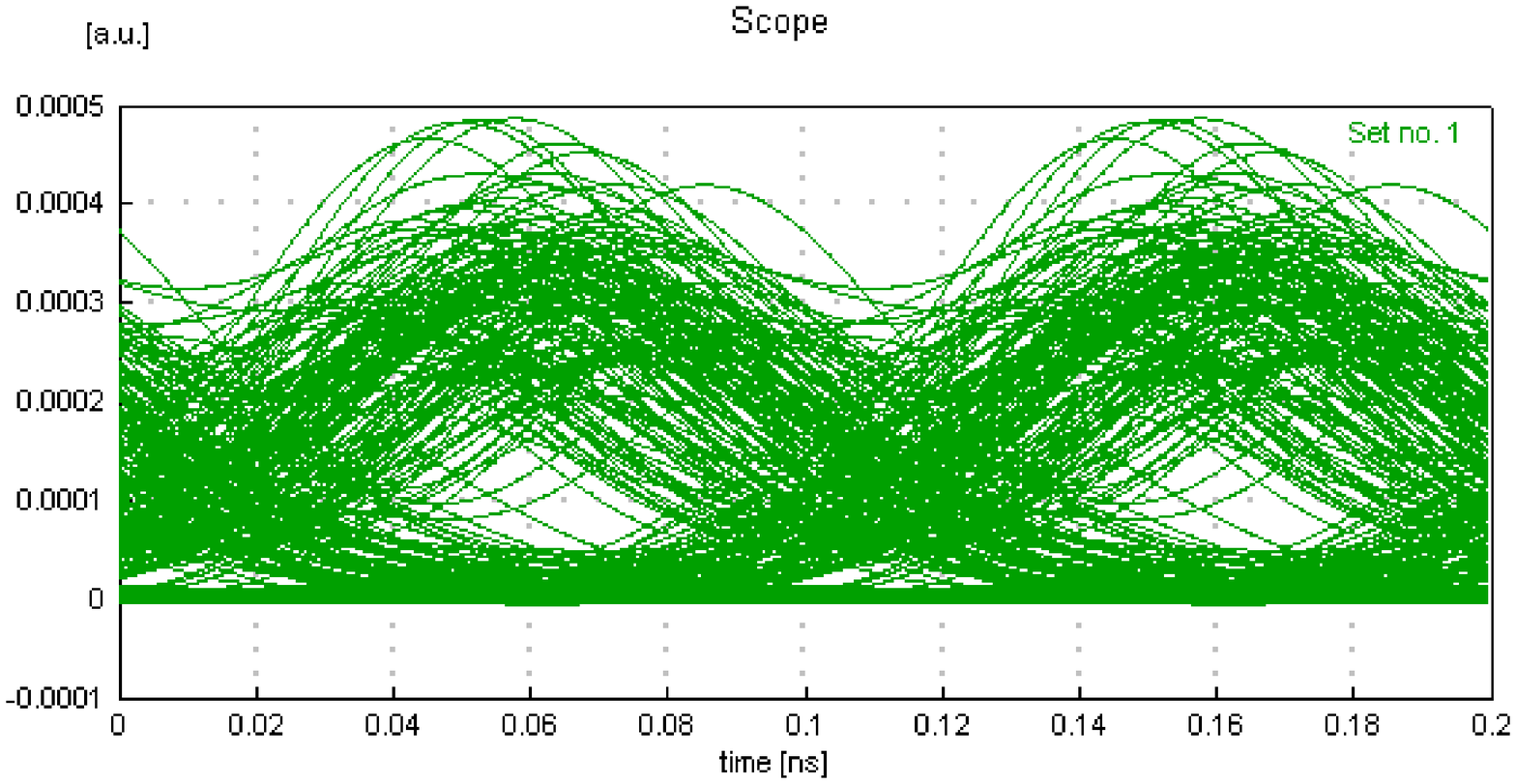}}
\end{minipage}
\begin{minipage}{78mm}
\scalebox{.37}{\includegraphics{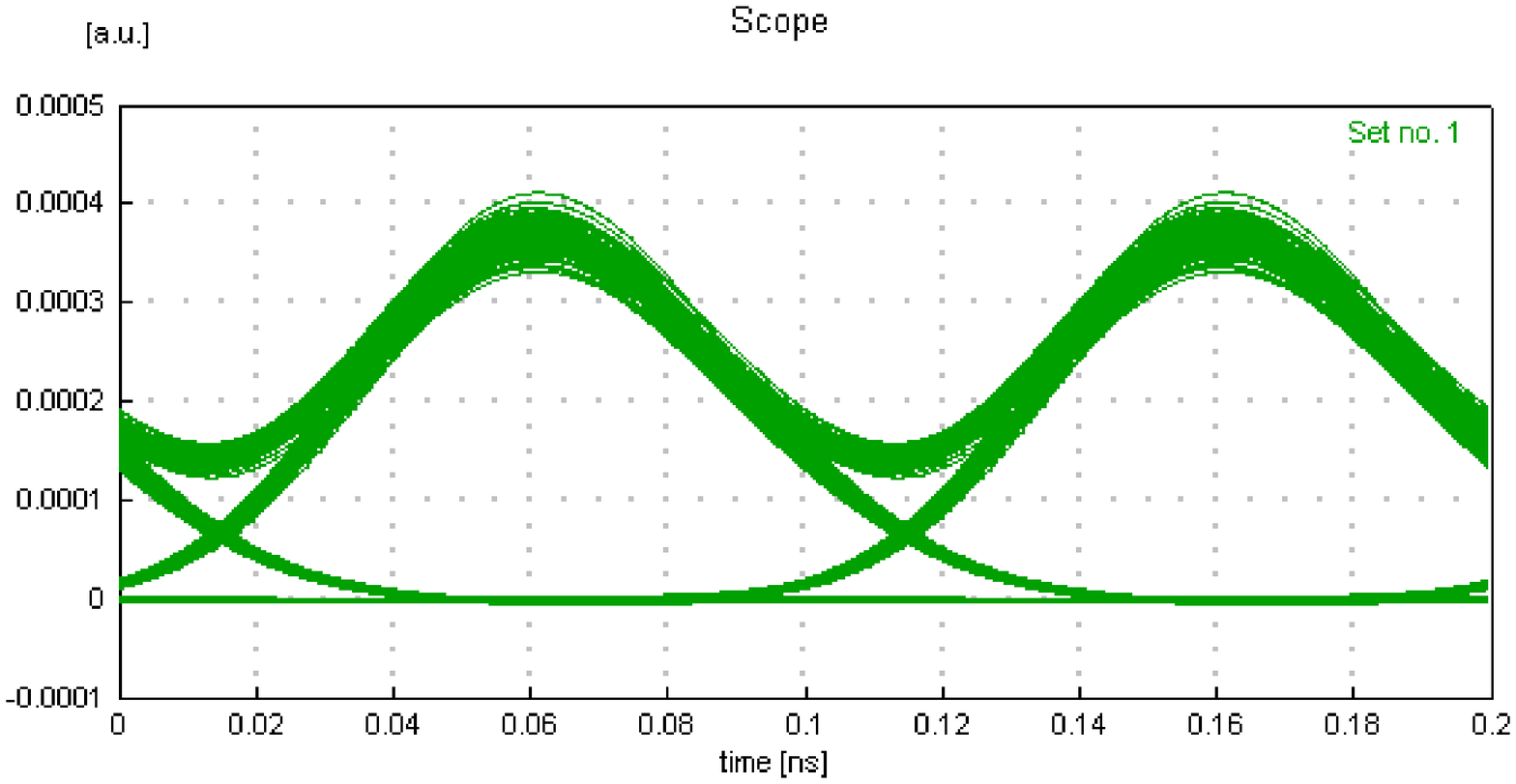}}
\end{minipage}
\end{center}
\caption{\label{simu4really}Received signals at the end of a
simulated link. Left: only the dispersion of the transmission
fiber is pre-compensated by the specialty fiber. Right: both
dispersion and nonlinearity are pre-compensated.}
\end{figure}

Using a commercial software, computer simulation has been carried
out to test the proposed method of nonlinearity compensation. As
in Fig.\ref{mirror}, the simulated link consists of a specialty
fiber, an OPC, and a transmission fiber that is of the negative
nonzero dispersion-shifted type, $200$ km long, with loss
coefficient $\alpha=0.2$ dB/km, dispersion $D=-8$ ps/nm/km,
dispersion slope $S=0.08$ ps/nm$^2$/km, effective mode area
$A_{\rm eff}=50$ $\mu$m$^2$, Kerr and Raman coefficients that are
typical of silica glass. The specialty fiber is a dispersion
compensating fiber of the same material, but with parameters
$(\alpha',D',S')=20\times(-\alpha,D,-S)$ and $A'_{\rm eff}=12.5$
$\mu$m$^2$. The nonlinearity of the specialty fiber can be
switched on and off. ASE noise is added at the two ends of the
link. The input consists of four WDM channels at $100$ GHz
spacing, all RZ modulated at $10$ Gb/s with $33\%$ duty. The power
of all optical pulses is peaked at $100$ mW when entering the
transmission fiber. Fig.\ref{simu4really} shows the received
signals without and with nonlinearity in the specialty fiber
respectively. Showing no nonlinear degradation, only the effect of
ASE noise, the graph on the right side demonstrates clearly the
compensation of optical nonlinearity.

\end{document}